# Accelerating Chemical Exchange Saturation Transfer Imaging Using a Model-based Deep Neural Network With Synthetic Training Data

Jianping Xu[1], Tao Zu[1], Yi-Cheng Hsu[2], Xiaoli Wang[3], Kannie W. Y. Chan[4], Yi Zhang[1]

[1]Key Laboratory for Biomedical Engineering of Ministry of Education, Department of Biomedical Engineering, College of Biomedical Engineering & Instrument Science, Zhejiang University, Hangzhou, Zhejiang, China

[2]MR Collaboration, Siemens Healthcare Ltd., Shanghai, China

[3]School of Medical Imaging, Weifang Medical University, Weifang, Shandong, China

[4]Department of Biomedical Engineering, City University of Hong Kong, Hong Kong, China

Address for correspondence:

Yi Zhang, PhD, Room 322, Zhou Yiqing Building, Yuquan Campus, Zhejiang University, 38 Zheda Road, Hangzhou 310027, China

Email: yizhangzju@zju.edu.cn

# ABSTRACT

**Purpose:** To develop a model-based deep neural network for high-quality image reconstruction of undersampled multi-coil chemical exchange saturation transfer (CEST) data.

**Theory and Methods:** Inspired by the variational network, the CEST image reconstruction equation is unrolled into a deep neural network (CEST-VN) with a *k*-space data-sharing block that takes advantage of the inherent redundancy in adjacent CEST frames and 3D spatial-frequential convolution kernels that exploit correlations in the *x*-$\omega$ domain. Additionally, a new pipeline based on multiple-pool Bloch-McConnell simulations is devised to synthesize multi-coil CEST data from publicly available anatomical MRI data. The proposed neural network is trained on simulated data with a CEST-specific loss function that jointly measures the structural and CEST contrast. The performance of CEST-VN was evaluated on three healthy volunteers and five brain tumor patients using retrospectively undersampled data with various acceleration factors, and compared with other state-of-the-art reconstruction methods.

**Results:** The proposed CEST-VN method generated high-quality CEST source images and APT-weighted (APTw) maps in healthy and brain tumor subjects, consistently outperforming GRAPPA, blind compressed sensing, and the original variational network. With the acceleration factors increasing from 3 to 6, CEST-VN with the same hyperparameters yielded similar and accurate reconstruction without apparent loss of details or increase of artifacts. The ablation studies confirmed the effectiveness of the joint CEST-specific loss function and data-sharing block used.

**Conclusions:** The proposed CEST-VN method can offer high-quality CEST source images and APTw maps from highly undersampled multi-coil data by integrating the deep-learning prior and multi-coil sensitivity encoding model.

# 1. INTRODUCTION

Chemical exchange saturation transfer (CEST) magnetic resonance (MR) imaging has proven to be a powerful and promising technique that can sensitively detect a wide range of biomolecules (1-4) in various diseases such as cancers (5-7) and epilepsies (8-10). However, the widespread clinical adoption of this technique has been hampered by its prolonged scan time, due to multiple image frames acquired over a range of saturation frequency offsets (11) .

Several image reconstruction strategies have been developed to enable rapid CEST data acquisition, which recovers images from undersampled *k*-space measurements by exploiting redundancies in the data. One of the most successful techniques is parallel imaging (12-16), which exploits the redundancy provided by multiple receive coils and turns the ill-posed problem into an overdetermined or determined one. An alternative way is utilizing the sparse prior in the underlying image, which is the foundation of compressed sensing (CS) (17-20). Besides, more aggressive undersampling can be achieved by exploiting the correlation in the spatial-frequential ($x$-$\omega$) domain, such as sharing data among different saturation frames (21) or leveraging the sparsity in the $x$-$\omega$ domain (22,23).

Recently, deep learning demonstrated immense potential in the MR image reconstruction field (24-28). Thanks to the offline training with big data, these deep-learning-based methods provide superior and superfast reconstruction without cumbersome parameter-tuning and time-consuming iteration processes, which is unavoidable for constrained reconstruction approaches such as CS. The downside of this is that, a large number of good-quality training data is essential to guarantee neural networks' reconstruction accuracy and generalization performance. This is challenging for imaging technologies that have not been widely used in clinical routines, such as CEST imaging. Moreover, many early works trained and evaluated their networks on homogeneous data obtained from a few institutions, making them hard to generalize to various practical situations (29,30).

Following the successful applications of deep learning in a wide range of fields, learning-based approaches have received great interest in CEST image reconstruction. Specifically, Guo et al. (31) used a U-Net (32) to directly learn the mapping from undersampled source images to CEST contrast maps. Wang et al. (33) introduced a recurrent network architecture to reconstruct amide proton transfer weighted (APTw) maps with the aid of fully-sampled T$_2$-weighted images. Both of the prior works used single-coil data, although multi-coil parallel acquisition is a standard for clinical MR imaging. Here, a novel deep learning method based on the variational network (26) is presented for fast multi-coil CEST imaging, which is dubbed CEST-VN. The main contributions of this work are summarized as follows:

1) A pipeline based on numerical Bloch-McConnell simulations is proposed to generate multi-coil CEST data from publicly available anatomical MR images, which greatly alleviates the hurdle of limited large-scale CEST data available for network training.

2) A *k*-space data-sharing block, which takes advantage of the inherent redundancy in neighboring CEST frames, is integrated into the network.

3) 3D spatial-frequential convolution kernels are used in the model-based network to efficiently exploit correlations in the *x-ω* domain.

4) A CEST-specific loss function, which jointly measures the error of reconstructed source images and the error of estimated CEST contrast, is designed for network training.

We train our network on simulated multi-coil CEST data and evaluate the network on experimental data from three healthy volunteers and five brain tumor patients. To the best of our knowledge, this is the first work integrating deep-learning and parallel imaging approaches to accelerate CEST acquisition (34). Evaluation results demonstrate the proposed CEST-VN method can achieve accurate and reliable reconstruction, and outperform the state-of-the-art approaches noticeably. Moreover, our approach can efficiently reconstruct a CEST map within 15 s on a single graphics card, which is fast enough to meet the clinical requirement.

## 2. THEORY

### 2.1 Signal Formulation

In CEST imaging, the *k*-space signals corresponding to different image frames are often sequentially acquired with varying radiofrequency (RF) saturation frequency offsets. An undersampled CEST MRI measurement with multiple receive coils can be formulated as

$$y_i(\mathbf{k},\omega) = \int_\mathbf{r} s(\mathbf{r},\omega)c_i(\mathbf{r})e^{-j\mathbf{k}\cdot\mathbf{r}}d\mathbf{r} + o(\mathbf{k},\omega), \qquad [1]$$

where $y_i(\mathbf{k},\omega)$ is the obtained signal of the $i$-th ($i = 1,...,N_c$) coil at the $k$-space position $\mathbf{k}$ and the saturation frequency offset $\omega$, $s(\mathbf{r},\omega)$ is the saturated image value at the image-space position $\mathbf{r}$, $c_i(\mathbf{r})$ is the corresponding coil sensitivity, and $o(\mathbf{k},\omega)$ refers to additive Gaussian noise in the measurements.

Let $\mathbf{S} \in \mathbb{C}^{N \times N_\omega}$ denote the stack of vectorized CEST source images along the z-spectrum direction,

$$\mathbf{S} = \begin{bmatrix} s(\mathbf{r}_1,\omega_1), & \cdots, & s(\mathbf{r}_1,\omega_{N_\omega}) \\ \vdots & & \vdots \\ s(\mathbf{r}_N,\omega_1), & \cdots, & s(\mathbf{r}_N,\omega_{N_\omega}) \end{bmatrix} \in \mathbb{C}^{N \times N_\omega}, \qquad [2]$$

where $N$ is the total number of voxels in an individual saturated image, and $N_\omega$ is the number of saturation frequency offsets. The signal model in Equation [1] can then be described in the matrix notation,

$$\mathbf{Y} = \mathbf{ES} + \mathbf{O}, \qquad [3]$$

where $\mathbf{Y} \in \mathbb{C}^{N_k N_c \times N_\omega}$ denotes the undersampled multi-coil $k$-space matrix, $N_k$ is the number of sampled $k$-space points, $\mathbf{O} \in \mathbb{C}^{N_k N_c \times N_\omega}$ refers to the complex noise matrix, and the encoding operator $\mathbf{E} \in \mathbb{C}^{N_k N_c \times N}$ models the physics of MR acquisition. Specifically, there is

$$\mathbf{ES} = \left[(\mathbf{F}_u \mathbf{C}_1 \mathbf{S}),...,(\mathbf{F}_u \mathbf{C}_{N_c} \mathbf{S})\right] \in \mathbb{C}^{N_k N_c \times N_\omega}, \qquad [4]$$

in which $\mathbf{F}_u \in \mathbb{C}^{N_k \times N}$ is the undersampled Fourier transform operator, and $\mathbf{C}_i = diag(c_i) \in \mathbb{C}^{N \times N}$ refers to the $i$-th diagonalized coil sensitivity map.

Given that Eq. [3] can be an ill-posed problem when incomplete $k$-space data is acquired, we formulate the reconstruction of $\mathbf{S}$ from $\mathbf{Y}$ measurements in the $k$-$\omega$ dimension as the following regularized optimization problem:

$$\hat{\mathbf{S}} = \arg\min_{\mathbf{S}} \frac{1}{2}\|\mathbf{ES} - \mathbf{Y}\|_F^2 + \lambda R(\mathbf{S}), \qquad [5]$$

where $\|\cdot\|_F$ denotes the Frobenius matrix norm, $R(\cdot)$ is a regularization function and $\lambda$

controls the trade-off between the data consistency term and the regularization term. Furthermore, when $R$ is convex, the gradient descent (GD) algorithm (35) is a feasible approach to solve Eq. [5], yielding the following intermediate update:

$$\mathbf{S}^{k+1} = \mathbf{S}^k - \gamma^k \left[ \mathbf{E}^H \left( \mathbf{E}\mathbf{S}^k - \mathbf{Y} \right) + \lambda (\nabla_{\mathbf{S}^k} R) \right]. \qquad [6]$$

Here, $\mathbf{S}^k$ and $\gamma^k$ are the estimated source image and the step size at the $k^{th}$ ($k = 1,...,K$) iteration, respectively. $\mathbf{E}^H$ represents the conjugate transpose of $\mathbf{E}$, and $\nabla_{\mathbf{S}^k} R$ signifies the gradient of the regularization term.

## 2.2 Unrolling the Variational Network (VN)

The regularization term $R(\cdot)$ is typically designed to utilize the sparse prior of the image in a certain transform domain, which can be achieved by the wavelet transform (17,36) and/or total variation (37). However, these predefined regularization functions may result in problems of residual artifacts and over-smoothing, due to simple and inaccurate sparse modeling (38). Given the considerations above, unlike the classical CS methods with fixed regularization, we propose to use the learning-based regularization, specifically variational network (26), to recover CEST source images from the undersampled *k*-space data, which is formulated as

$$R(\mathbf{S}) = \sum_{i=1}^{N_v} \langle \Phi_i (\mathbf{D}_i \mathbf{S}), \mathbf{1} \rangle. \qquad [7]$$

Here, $N_v$ refers to the number of tunable terms in the regularization, $\langle \, , \, \rangle$ denotes the inner product, $\mathbf{D} : \mathbb{C}^{N \times N_\omega} \to \mathbb{C}^{N \times N_\omega}$ is a linear convolution operator achieved by a spatial-frequential kernel $D \in \mathbb{C}^{d \times d \times d}$ ($d$ refers to the size of the kernel), $\Phi : \mathbb{C}^{N \times N_\omega} \to \mathbb{C}^{N \times N_\omega}$ designates a non-linear potential function, and $\mathbf{1}$ refers to a vector of ones. Fundamentally, Eq. [7] can be regarded as a relaxation of the commonly used regularization, such as sparse regularization (17,26) and locally low-rank regularization (39,40). Substituting Eq. [7] into Eq. [6] yields

$$\mathbf{S}^{k+1} = \mathbf{S}^k - \gamma^k \mathbf{E}^H \left( \mathbf{E}\mathbf{S}^k - \mathbf{Y} \right) - \sum_{i=1}^{N_v} \mathbf{D}_i^T \varphi_i (\mathbf{D}_i \mathbf{S}^k), \qquad [8]$$

where $\mathbf{D}_i^T : \mathbb{C}^{N \times N_\omega} \to \mathbb{C}^{N \times N_\omega}$ denotes the convolution operator achieved by the transpose of $D_i$, and $\varphi_i : \mathbb{C}^{N \times N_\omega} \to \mathbb{C}^{N \times N_\omega}$ is the activation function defined by the first derivative of the potential function $\Phi_i$. Note that the weight of the regularization term in Eq. [8] is omitted since it can be implicitly included in tunable activation functions.

By unrolling $K$ update steps of the gradient descent process, the numerical optimization problem illustrated in Eq. [8] can be eventually converted into a variational network with $K$ iterative blocks, and the $k^{th}$ ($k=1,...,K$) block is illustrated in Fig. 1B. Specifically, to efficiently exploit correlations in the spatial-frequential ($x$-$\omega$; each CEST source image labeled by its saturation frequency offset) or spatial-temporal ($x$-$t$; each CEST source image labeled by its acquisition time) domain, the regularization module in each block contains $N_v$ sets of 3D spatial-frequential or spatial-temporal kernels, $D_i$ and $D_i^T$, that operate in the $x$-$\omega$ or $x$-$t$ space for a series of 2D CEST source images. Between each set of $D_i$ and $D_i^T$ is a tunable activation function $\varphi_i$, ensuring an efficient representation of sparse features in the convolutional space. To deal with the complex-valued input, all the kernels and activation functions in each iterative block contain two parallel channels. In contrast to the classical constrained reconstruction methods, the variational network relaxes the numerical iteration procedure by allowing numerous parameters learnable at each network layer, including kernel weights $\mathbf{W}_\mathbf{D}^k$, parametrized activation functions $\boldsymbol{\varphi}^k = \{\varphi_1^k,...\varphi_{N_v}^k\}$, and step size $\gamma^k$:

$$\boldsymbol{\theta}^k = \{\mathbf{W}_\mathbf{D}^k, \boldsymbol{\varphi}^k, \gamma^k\}. \qquad [9]$$

## 2.3 Proposed Model: CEST-VN

CEST-VN is a model-based network that reconstructs the CEST source image from the undersampled, multi-coil *k*-space data. Fig. 1 schematically illustrates the architecture of CEST-VN, which consists of two important components: 1) a *k*-space data sharing block that takes advantage of adjacent frames in the frequency offset dimension, and 2) an image reconstruction network with a trainable regularization module and a model-based data consistency module in each iterative block. These two structures are concatenated successively, and the former provides the reconstruction network with an improved estimation as initial input.

**1) K-space Data Sharing (DS) Block**

In order to detect the CEST effect reliably or for quantitative CEST imaging, the frequency distribution of RF saturation pulses is often required to be relatively dense over a z-spectral range, especially around the frequencies of interest. As the CEST effects caused by adjacent saturation frequencies are similar in practice, it is reasonable to consider that the *k*-space samples from neighboring frames capture similar information. The local similarity or correlation in *k-ω* space can be introduced as a strong prior contributing to the reconstruction from highly undersampled *k*-space data (21,27,41,42). Inspired by these earlier works, a DS block performing in the *k-ω* domain is presented. As detailed in Fig. 1A, the missing multi-coil *k*-space samples in the current CEST frame $\mathbf{y}_\omega$ are filled using the sampled data from two adjacent frames $\mathbf{y}_{\omega-1}$ and $\mathbf{y}_{\omega+1}$ at the corresponding positions, which yields a more complete *k*-space. But rather than considering the shared samples as the final estimation like the keyhole method (21), the output of the DS block will be fed into the subsequent image reconstruction network for further optimization. Compared to the previous efforts (27) that iteratively combine the adjacent undersampled *k*-space of dynamic MR images, the sharing strategy in our approach has two main differences: 1) the DS block in CEST-VN only provides an initial estimation and is independent of the succeeding image reconstruction network, and 2) the sharing is only allowed from the closest frames instead of over a larger range. Both strategies aim to avoid the loss of distinction among different CEST frames, which is crucial to ultimate CEST analysis.

**2) Image Reconstruction Network**

Following the *k*-space DS block is an unrolled variational network with $K=10$ iterative blocks. In detail, the regularization module in each iterative block contains $N_v=36$ sets of 3D kernels (size=7×7×7) and corresponding activation functions, each of which is fitted by a weighted combination of $N_f=31$ Gaussian radial basis functions (26). In addition, the pre-computed coil sensitivity maps—$\mathbf{C}$, along with the undersampling masks, are simultaneously sent into each iterative block and used in the encoding operators—$\mathbf{E}$ and $\mathbf{E}^H$—inside the model-based data consistency module. Given the trainable parameter set in each iterative block depicted in Eq. [9], the source image reconstruction process achieved by CEST-VN can be described as a mapping function of the undersampled *k*-space data and network parameter:

$$\mathbf{S} = f(\mathbf{Y}|\Theta), \qquad [10]$$

where $\Theta = \{\theta^1, \theta^2, ..., \theta^K\}$ is a set of all the trainable parameters, and $f: \mathbb{C}^{N_k N_c \times N_\omega} \to \mathbb{C}^{N \times N_\omega}$ denotes the mapping learned by the neural network.

**3) CEST-specific Loss Function**

A CEST-specific loss function—$L(\Theta)$—is designed to optimize the trainable network parameters, which jointly considers the error of reconstructed source images and the error of estimated CEST contrast as

$$L(\Theta) = \sum_{\{\mathbf{Y},\mathbf{S}^*\} \in \Omega} \|f(\mathbf{Y}|\Theta) - \mathbf{S}^*\|_2^2 + \mu \sum_{\{\mathbf{Y},\mathbf{S}^*\} \in \Omega} \|\Lambda(f(\mathbf{Y}|\Theta)) - \Lambda(\mathbf{S}^*)\|_2^2, \qquad [11]$$

where $\|\cdot\|_2$ represents the L$_2$ norm, $\mathbf{S}^*$ denotes the ground-truth source images, $\Omega$ refers to the training dataset, and $\Lambda: \mathbb{C}^{N \times N_\omega} \to \mathbb{C}^{N \times N_\Lambda}$ means the operator used to generate $N_\Lambda$ CEST contrast maps. Here, following the magnetization transfer ratio asymmetry (MTR$_{asym}$) analysis method, $\Lambda(\mathbf{S}) \in \mathbb{C}^{N \times N_\Lambda}$ is stacked by $N_\Lambda$ column vectors, the $j$-th ($j = 1, ..., N_\Lambda$) of which is defined by $\mathbf{S}(-\omega_j) - \mathbf{S}(+\omega_j)$ to quantify the CEST effect at the frequency offset of $\omega_j$. The first term in Eq. [11] enforces the concordance of reconstructed source images with the reference ones, mostly preserving the structural information. And the second term ensures the consistency of the CEST effect with the ground-truth, mainly retaining molecular information. The trade-off between the two terms is controlled by $\mu$ empirically chosen as 10 in this work.

## 3. METHODS

### 3.1 CEST K-space Data Synthesis

Because of limited experimental CEST *k*-space data available and to strengthen the generalizability of the neural network, a pipeline based on the numerical simulation was proposed to synthesize multi-coil CEST *k*-space data from the publicly available fastMRI brain dataset (43,44), which consists of large-scale multi-coil data from anatomical T$_1$-weighted and T$_2$-weighted sequences. Different from the anatomical images, the CEST data has an extra dimension with the signal of each voxel modulated by the underlying z-spectrum as

$$s(\mathbf{r},\omega) = s_0(\mathbf{r})\varsigma(\mathbf{r},\omega), \qquad [12]$$

where $s(\mathbf{r},\omega)$ is the saturated complex-valued signal specified by the position $\mathbf{r}$ and the saturation frequency $\omega$, $s_0(\mathbf{r})$ is the unsaturated complex signal at the position $\mathbf{r}$, and $\varsigma(\mathbf{r},\omega) \in \mathbb{R}^{N_\omega}$ is the corresponding magnitude z-spectrum. By treating the fastMRI anatomical image as $s_0(\mathbf{r})$, we can generate the CEST source images via the modulation of simulated z-spectra as in Eq. [12]. The procedures of CEST brain data generation are shown schematically in Fig. 2.

**Z-spectrum Simulation and Data Pre-processing**

The process of CEST MR imaging can be well described by the Bloch-McConnell equation (45). Here, a three-pool Bloch-McConnell model, including water, amide, and symmetric semisolid magnetization transfer (MT) pools, was carried out to generate the z-spectra. The same saturation parameters with the following in vivo experimental setup were used in the Bloch-McConnell simulations. A total of 61,200 z-spectra within -6 to 6 ppm were generated for a range of relaxation times and CEST pool concentrations, with detailed simulation parameters listed in Supporting Information Table S1. As the fastMRI repository consists of data with different numbers of coil channels, only data with more than 16 channels was used and then channel-compressed (46) to 16 virtual coils. For each image volume, the central 10 slices were extracted, each of which was subsequently cropped and resized to a matrix size of 96 × 96 in the image domain. The pre-processed 2D anatomical images were regarded as $s_0(\mathbf{r})$ of Eq. [12].

**Multi-coil CEST Data Generation**

Firstly, the anatomical brain image was segmented into tumor and non-tumor regions. For the fastMRI brain images with suspected tumors, a clustering algorithm based on k-means (47,48) was used to identify and segment the tumors in each anatomical image. As for the anatomical brain images with no obvious abnormality, curved irregular areas were randomly synthesized as tumor regions among half of the subjects, introducing more randomness to the training set. Secondly, in order to mimic the distribution of amide protons in tumor tissues and normal tissues, soft textures were obtained from a large number of natural-scene images from ImageNet (49) through smoothing, which were then randomly incorporated into the tumor and non-tumor regions, respectively.

Thirdly, a mapping between the solute concentrations in the amide pool and the gray-scale image values was pre-defined, based on which each voxel was matched with a z-spectrum in the aforementioned 61,200 simulated ones, yielding an $x$-$\omega$ tensor, i.e., z-map. Fourthly, by modulating a multi-coil anatomical MR image in the $x$-$\omega$ domain with the corresponding z-map, we obtained a cluster of simulated multi-coil saturated images and their $k$-space data, which were used as the training samples.

Following the simulation procedures above, we obtained 9250 samples using MATLAB R2020b (MathWorks, Natick, MA, USA). Examples of saturated images and associated APTw maps can be found in Supporting Information Fig. S1. For each sample, coil sensitivity maps were calculated from the center $k$-space data with a size of 16 × 16 using ESPIRiT (50), and the image reconstructed by SENSE (12) from full $k$-space data was regarded as the training ground-truth. Additionally, both the undersampled $k$-space data and reference images were normalized according to the maximum absolute value of the reconstructed image from zero-filled undersampled $k$-space data.

## 3.2 In Vivo Data Acquisition

To evaluate the performance of the proposed approach on experimental data, full $k$-space CEST data of three healthy volunteers and five brain tumor patients were acquired using a 3T MRI scanner (MAGNETOM Prisma, Siemens Healthcare, Erlangen, Germany). All human studies were approved by the local institutional review board, and informed consent was obtained from each subject.

For each subject, a frequency-stabilized CEST sequence was performed with single-slice turbo-spin-echo readout (51,52) using the following acquisition parameters: flip angle = 90°, echo time = 6.7 ms, repetition time (TR) = 3000 ms, slice thickness = 5 mm, field of view = 212 × 186 mm$^2$, acquisition resolution = 2.2 × 2.2 mm$^2$, and turbo factor = 96. Specifically, the CEST saturation was achieved using a train of ten Gaussian pulses, each with a 100-ms duration and 2-μT saturation power, separated by an interval of 10 ms (53). Furthermore, 54 CEST frames were acquired, including one unsaturated image and 53 saturated images within -6 to 6 ppm. Moreover, for $B_0$ inhomogeneity correction, the WASSR method (54) was used with TR = 2000 ms, saturation power = 0.5 μT, saturation duration = 200 ms, and 26 saturation frequencies from -1.5 to 1.5 ppm in 0.125 ppm steps, while the other scan parameters were the same as the aforementioned CEST sequence. The coil compression, sensitivity map calculation and data normalization processes were the same as those for simulated datasets.

## 3.3 Network Training

CEST-VN was trained only with the simulated multi-coil data and tested using experimental data acquired from healthy volunteers and tumor patients. Regarding training data, the 9250 simulated samples were randomly divided into two subsets, i.e. the training dataset (85%) and the validation dataset (15%), while the latter was used to tune the network hyperparameters. Cartesian 1D variable-density sampling masks were used to retrospectively undersample the multi-coil *k*-space data in the phase-encoding direction in the *k-ω* space. The reduction factors were R = 3, 4, 5, and 6. Specifically, all the networks were trained for 50 epochs using the adaptive moment estimation (Adam) optimizer (55) (momentum $\beta_1$ = 0.9, $\beta_2$ = 0.999, batch size of 4, initial learning rate of $10^{-4}$), with a learning rate rescheduled every 20 epochs by a multiplicative factor γlr = 0.5. The trainable parameters of step size $\gamma^k$ were initialized as 1. Both the training and testing pipelines were implemented using the Pytorch framework (56) on 4 NVIDIA 2080Ti GPUs with CUDA and cuDNN support. All the source code is available from the authors upon reasonable request. The deployment of the trained CEST-VN was able to be completed within 15 s on a single graphics card.

## 3.4 Evaluation

**1) CEST Analysis and Evaluation Metrics**

After $B_0$ inhomogeneity correction via the WASSR method, the $MTR_{asym}$ analysis was used to quantify the CEST effect as

$$MTR_{asym}(-\omega) = [s(-\omega) - s(\omega)] / s_0, \qquad [13]$$

where $\omega$ is the saturation frequency, s is the saturated CEST signal, and $s_0$ is the unsaturated signal. Specifically, the APTw maps were calculated by setting $\omega$ = 3.5 ppm. To quantitatively evaluate the performance of the proposed method, fully-sampled CEST source images and APTw maps were treated as the reference ($\mathbf{x}^*$). For the reconstructed magnitude source images and APTw maps from undersampled *k*-space data ($\mathbf{x}$), we calculated the normalized root mean square error (nRMSE) and the peak signal-to-noise ratio (PSNR) against the ground-truth as follows:

$$\mathrm{nRMSE}(\mathbf{x}, \mathbf{x}^*) = \frac{1}{\max(\mathbf{x}^*) - \min(\mathbf{x}^*)} \sqrt{\frac{\sum_i^N |x_i - x_i^*|^2}{N}}, \quad [14]$$

$$\mathrm{PSNR}(\mathbf{x}, \mathbf{x}^*) = 10 \log_{10} \frac{N \cdot \max(\mathbf{x}^*)^2}{\|\mathbf{x} - \mathbf{x}^*\|_2^2}, \quad [15]$$

where $x_i$ denotes the *i*-th ($i = 1, ..., N$) voxel of $\mathbf{x}$, and $x_i^*$ denotes the *i*-th voxel of $\mathbf{x}^*$. Additionally, the mean absolute error (MAE) was used to evaluate the quality of the reconstructed z-spectrum and $\mathrm{MTR}_{\mathrm{asym}}$ spectrum:

$$\mathrm{MAE}(\mathbf{x}, \mathbf{x}^*) = \frac{1}{N} \|\mathbf{x} - \mathbf{x}^*\|_1, \quad [16]$$

where $\|\cdot\|_1$ is the $L_1$ norm.

**2) Comparison Studies**

We compared the proposed CEST-VN method with three state-of-the-art approaches, i.e. GRAPPA (13), blind compressed sensing (BCS) (57,58), and the original VN (26). All these methods were carefully adjusted to optimally suit the current experiment setup to promote a fair comparison. For BCS, the dictionary consisted of 40 temporal basis functions, which were used to compose the signal evolution of each pixel, as described in Ref. (57). For VN, the original open-source implementation was used (26) to reconstruct the CEST source image frame by frame, and the training was executed with the conventional mean squared error loss function as described in Ref. (26). The other training procedures of VN were consistent with those of CEST-VN.

**3) Effects of Acceleration Factors and Ablation Studies**

To evaluate the reconstruction performance of the proposed network for different acceleration factors, we tested CEST-VN using data with retrospective undersampling factors from 3 to 6. Furthermore, in order to demonstrate the effectiveness of the proposed CEST-specific loss function $L(\Theta)$ in Eq. [11], the network was additionally trained with two different loss functions: 1) only the first term of Eq. [11], namely source image loss, and 2) only the second term of Eq. [11], namely CEST contrast loss. Also, a retrospective ablation study was carried out without the *k*-space DS block to illustrate the efficacy of the *k*-space data-sharing procedure.

# 4. RESULTS

## 4.1 Comparison Studies

The reconstruction results from various methods were first compared using experimental data from healthy volunteers, as shown in Figs. 3-4. As for the APTw maps, Fig. 3 shows the proposed CEST-VN method yielded a better agreement with the fully-sampled reference and a smaller reconstruction error than the GRAPPA, BCS, and VN approaches using the same undersampling pattern (R = 6). Specifically, the nRMSE of CEST-VN was 61.9%, 15.9% and 44.4% lower than that of GRAPPA, BCS and original VN, respectively. As for the source CEST images, Fig. 4 shows the same pattern as that in color-coded APTw maps, with CEST-VN generating the best image quality. At various saturation frequency offsets, CEST-VN consistently outperformed the other state-of-the-art methods (GRAPPA, BCS and VN). Notably, the fine image structures (red arrows), such as the brain midline or the sulcus, were best preserved by the proposed CEST-VN method.

To further validate CEST-VN on clinical data, evaluation experiments were conducted on five brain tumor patients. Figs. 5-7 compare the reconstruction and CEST analysis results from different methods on a newly-diagnosed glioma patient. Fig. 5 depicts the APTw maps from 4-fold-accelerated reconstruction, undersampling pattern used, and corresponding error maps versus the fully-sampled reference. Stronger residual aliasing artifacts were observed for GRAPPA reconstruction than the other methods, which was possibly because the GRAPPA method was not suited for the variably-undersampled *k*-space (Fig. 5B). Both BCS and VN generated error maps with non-random residues, especially in the tumor region for VN (Fig. 5C). In contrast, CEST-VN had superior performance in both artifact removal and detail retention, yielding a better agreement with the reference and a lower nRMSE value.

Fig. 6 displays reconstructed CEST source images, along with z-spectra and $MTR_{asym}$ spectra from a chosen region of interest in the tumor (indicated by the red circle). The z-spectra and $MTR_{asym}$ spectra averaged over the whole tumor region can be found in Supporting Information Fig. S2. Among all reconstruction methods tested, CEST-VN produced the highest PSNR in the source image (Fig. 6A), smallest nRMSE in the error map (Fig. 6B), and lowest MAE in z-spectra (Fig. 6C) and $MTR_{asym}$ spectra (Fig. 6D). Notably, although BCS yielded similar quality of reconstructed source images to CEST-VN in terms of visual inspection and quantitative errors, its $MTR_{asym}$ spectrum substantially deviated away from the reference one.

Fig. 7 shows the voxel-wise correlation and Bland-Altman analyses between accelerated and fully-sampled $MTR_{asym}$ values in the whole tumor region of the patient shown in Figs. 5-6. As for the correlation analysis (Fig. 7A), $MTR_{asym}$ values obtained by CEST-VN more closely correlated with the reference ones with a coefficient of determination $R^2 = 0.95$, while the other methods had $R^2 \leq 0.82$. As for the Bland-Altman analysis, the difference from the ground truth was mostly (2 times the standard deviation) within a range of 0.95% for CEST-VN, whereas the other methods had difference ranges $\geq 1.94\%$.

Fig. 8 illustrates the pooled errors from different reconstruction methods on all five brain tumor patients with varying acceleration factors from 3 to 6. The numeric mean (± standard deviation) nRMSE values of APTw maps and CEST source images can be found in Supporting Information Tables S2 and S3, respectively. It is evident that CEST-VN generated the smallest errors for both the APTw maps (Fig. 8A) and CEST source images (Fig. 8B) among all methods across all acceleration factors.

## 4.2 Effects of Acceleration Factors and Ablation Studies

Fig. 9 displays reconstructed APTw maps with different acceleration factors from the proposed CEST-VN method. The CEST-VN settings yielded similar and high-quality APTw maps without apparent loss of details or significant increase of artifacts for various acceleration factors, indicating its good robustness. Specifically, the nRMSE increased from 1.59% to 2.52% as the acceleration factor increased from 3 to 6.

Fig. 10 illustrates the effectiveness of the proposed CEST-specific loss function and the *k*-space DS block. The network trained using only the source image loss function resulted in obvious residual noise in the APTw map, although the details were well recovered (Fig. 10B). On the contrary, the network trained using only the CEST contrast loss function had a high visual signal-to-noise ratio, but an obvious over-smoothing problem occurred (Fig. 10C). By including both the source image and CEST contrast loss functions, APTw maps approximating the fully-sampled reference were obtained, with a good balance of image details and smoothness (Figs. 10D-E). Moreover, the proposed *k*-space DS strategy helped reduce reconstruction errors with nRMSE of 0.64% vs. 0.68% in APTw maps (Fig. 10G) and MAE of 0.103 vs. 0.112 in $MTR_{asym}$ spectra (Fig. 10H).

## 5. DISCUSSION

In this work, we further developed the variational network (26) and introduced the first deep neural network for multi-coil CEST image reconstruction. Given that multiple images are sequentially acquired with varying RF saturation offsets in CEST imaging, rich information is captured in the obtained CEST source images, including both the structural contrast in spatial space and molecular information in the saturation frequency dimension. Hence, instead of implementing the reconstruction in a frame-by-frame manner, CEST source images at different saturation offsets were regarded as a whole and 3D spatial-frequential kernels were used to extract features in the entire $x$-$\omega$ space. Meanwhile, a $k$-space data-sharing block was integrated into our network to take advantage of the inherent redundancy in adjacent CEST frames. These modifications enabled the CEST-VN architecture to better accommodate the image contrast of CEST data, resulting in superior performance on highly undersampled data to existing methods (GRAPPA, BCS and original VN) as in Figs. 3-8.

The performance of a reconstruction network is highly dependent on the loss function (38,59). Our experiments indicated the original loss function of VN that measures the structural contrast in source images might yield inaccurate APTw maps with noise-like artifacts, as shown in Fig. 10B. Hence, to accommodate the characteristics of CEST images, a CEST-specific loss function was designed for network training, which jointly measures the structural image information and molecular CEST effect. The experimental results proved that the network trained with the CEST-specific loss function was able to generate CEST source images with retained structural details (Fig. 4), APTw maps with accurate molecular contrast (Fig. 5), and reliable z-spectra and $MTR_{asym}$ spectra (Fig. 6). In this work, we only adopted $L_2$-norm loss functions that were efficient to compute. However, alternative image metrics can be used for loss function design, such as the structural similarity index (SSIM) (60), which considers local patch statistics. Thus, a further improvement of our network may be realized by using a combined loss function with different image metrics.

Sufficient and high-quality training data is a critical prerequisite for successful network learning, which, however, is hardly available for new imaging technologies, such as CEST imaging. To address this "small data" issue, we presented a novel strategy that generated multi-coil CEST brain data from public $T_1$-weighted and $T_2$-weighted datasets (fastMRI). Specifically, z-spectra were modulated into 2D anatomical images to create 3D spatial-frequential CEST data, which were simulated using a multiple-pool Bloch-McConnell model. In addition, natural-scene images from ImageNet were used to create tumor-like textures in the simulated CEST data. Importantly, the proposed CEST-VN model was trained only with the synthesized CEST data without using any experimental CEST data. However, the proposed network performed well on actual CEST data in both healthy volunteers and brain tumor patients, proving the excellent generalizability of our network and the effectiveness of the proposed method for simulating multi-coil CEST training data. In future investigations, we will evaluate the generalization of our network on more diseases such as stroke and epilepsy.

## 6. CONCLUSION

Inspired by the variational network, a novel deep neural network with a *k*-space data-sharing block and 3D spatial-frequential convolution kernels is proposed for fast CEST imaging dubbed CEST-VN. The CEST-VN is trained with only simulated multi-coil CEST data and a CEST-specific loss function. Experimental results indicate that the proposed CEST-VN approach generalizes well to experimental CEST data with various acceleration factors on healthy volunteers and brain tumor patients, outperforming conventional approaches and the original variational network.

## ACKNOWLEDGMENTS

National Natural Science Foundation of China: 81971605. Key R&D Program of Zhejiang Province: 2022C04031. Leading Innovation and Entrepreneurship Team of Zhejiang Province: 2020R01003. This work was supported by the MOE Frontier Science Center for Brain Science & Brain-Machine Integration, Zhejiang University.

## DATA AVAILABILITY STATEMENT

The CEST-VN code that supports the findings of this study is openly available from the authors upon reasonable request.

# REFERENCES

1. Ward KM, Aletras AH, Balaban RS. A new class of contrast agents for MRI based on proton chemical exchange dependent saturation transfer (CEST). Journal of Magnetic Resonance 2000;143(1):79-87.
2. van Zijl PCM, Lam WW, Xu J, Knutsson L, Stanisz GJ. Magnetization Transfer Contrast and Chemical Exchange Saturation Transfer MRI. Features and analysis of the field-dependent saturation spectrum. Neuroimage 2018;168:222-241.
3. Jones KM, Pollard AC, Pagel MD. Clinical applications of chemical exchange saturation transfer (CEST) MRI. Journal of Magnetic Resonance Imaging 2018;47(1):11-27.
4. Zhou J, Zaiss M, Knutsson L, Sun PZ, Ahn SS, Aime S, Bachert P, Blakeley JO, Cai K, Chappell MA, Chen M, Gochberg DF, Goerke S, Heo HY, Jiang S, Jin T, Kim SG, Laterra J, Paech D, Pagel MD, Park JE, Reddy R, Sakata A, Sartoretti-Schefer S, Sherry AD, Smith SA, Stanisz GJ, Sundgren PC, Togao O, Vandsburger M, Wen Z, Wu Y, Zhang Y, Zhu W, Zu Z, van Zijl PCM. Review and consensus recommendations on clinical APT-weighted imaging approaches at 3T: Application to brain tumors. Magnetic resonance in medicine 2022;88(2):546-574.
5. Zhang H, Yong X, Ma X, Zhao J, Shen Z, Chen X, Tian F, Chen W, Wu D, Zhang Y. Differentiation of low- and high-grade pediatric gliomas with amide proton transfer imaging: added value beyond quantitative relaxation times. European Radiology 2021;31(12):9110-9119.
6. Jia X, Wang W, Liang J, Ma X, Chen W, Wu D, Lai C, Zhang Y. Risk stratification of abdominal tumors in children with amide proton transfer imaging. European Radiology 2022;32(4):2158-2167.
7. Jiang S, Eberhart CG, Lim M, Heo HY, Zhang Y, Blair L, Wen Z, Holdhoff M, Lin D, Huang P, Qin H, Quinones-Hinojosa A, Weingart JD, Barker PB, Pomper MG, Laterra J, van Zijl PCM, Blakeley JO, Zhou J. Identifying Recurrent Malignant Glioma after Treatment Using Amide Proton Transfer-Weighted MR Imaging: A Validation Study with Image-Guided Stereotactic Biopsy. Clinical Cancer Research 2019;25(2):552-561.
8. Wen Q, Wang K, Hsu YC, Xu Y, Sun Y, Wu D, Zhang Y. Chemical exchange saturation transfer imaging for epilepsy secondary to tuberous sclerosis complex at 3 T: Optimization and analysis. NMR in Biomedicine 2021;34(9):e4563.
9. Wang K, Wen Q, Wu D, Hsu YC, Heo HY, Wang W, Sun Y, Ma Y, Wu D, Zhang Y. Lateralization of temporal lobe epileptic foci with automated chemical exchange saturation transfer measurements at 3 Tesla. EBioMedicine 2023;89:104460.
10. Davis KA, Nanga RP, Das S, Chen SH, Hadar PN, Pollard JR, Lucas TH, Shinohara RT, Litt B, Hariharan H, Elliott MA, Detre JA, Reddy R. Glutamate imaging (GluCEST) lateralizes epileptic foci in nonlesional temporal lobe epilepsy. Science Translational Medicine 2015;7(309):309ra161.
11. Zhang Y, Zu T, Liu R, Zhou J. Acquisition sequences and reconstruction methods for fast chemical exchange saturation transfer imaging. NMR in Biomedicine 2022:e4699.
12. Pruessmann KP, Weiger M, Scheidegger MB, Boesiger P. SENSE: sensitivity encoding for fast MRI. Magnetic Resonance in Medicine 1999;42(5):952-962.
13. Griswold MA, Jakob PM, Heidemann RM, Nittka M, Jellus V, Wang JM, Kiefer B, Haase A. Generalized Autocalibrating Partially Parallel Acquisitions (GRAPPA). Magnetic Resonance in Medicine 2002;47(6):1202-1210.
14. Zhang Y, Heo HY, Lee DH, Jiang S, Zhao X, Bottomley PA, Zhou J. Chemical exchange saturation transfer (CEST) imaging with fast variably‑accelerated sensitivity encoding (vSENSE). Magnetic resonance in medicine 2017;77(6):2225-2238.

15. Zhang Y, Heo HY, Jiang S, Zhou J, Bottomley PA. Fast 3D chemical exchange saturation transfer imaging with variably‐accelerated sensitivity encoding (vSENSE). Magnetic resonance in medicine 2019;82(6):2046-2061.
16. Zu T, Sun Y, Wu D, Zhang Y. Joint K‐space and Image‐space Parallel Imaging (KIPI) for accelerated chemical exchange saturation transfer acquisition. Magnetic Resonance in Medicine 2023;89(3):922-936.
17. Lustig M, Donoho D, Pauly JM. Sparse MRI: The application of compressed sensing for rapid MR imaging. Magnetic Resonance in Medicine: An Official Journal of the International Society for Magnetic Resonance in Medicine 2007;58(6):1182-1195.
18. Heo HY, Xu X, Jiang S, Zhao Y, Keupp J, Redmond KJ, Laterra J, van Zijl PCM, Zhou J. Prospective acceleration of parallel RF transmission-based 3D chemical exchange saturation transfer imaging with compressed sensing. Magnetic resonance in medicine 2019;82(5):1812-1821.
19. Heo HY, Zhang Y, Lee DH, Jiang S, Zhao X, Zhou J. Accelerating chemical exchange saturation transfer (CEST) MRI by combining compressed sensing and sensitivity encoding techniques. Magnetic resonance in medicine 2017;77(2):779-786.
20. She H, Greer JS, Zhang S, Li B, Keupp J, Madhuranthakam AJ, Dimitrov IE, Lenkinski RE, Vinogradov E. Accelerating chemical exchange saturation transfer MRI with parallel blind compressed sensing. Magnetic resonance in medicine 2019;81(1):504-513.
21. Varma G, Lenkinski R, Vinogradov E. Keyhole chemical exchange saturation transfer. Magnetic Resonance in Medicine 2012;68(4):1228-1233.
22. Lee H, Chung JJ, Lee J, Kim S-G, Han J-H, Park J. Model-based chemical exchange saturation transfer MRI for robust z-spectrum analysis. IEEE Transactions on Medical Imaging 2019;39(2):283-293.
23. Kwiatkowski G, Kozerke S. Accelerating CEST MRI in the mouse brain at 9.4 T by exploiting sparsity in the Z‐spectrum domain. NMR in Biomedicine 2020;33(9):e4360.
24. Wang S, Su Z, Ying L, Peng X, Zhu S, Liang F, Feng D, Liang D. Accelerating magnetic resonance imaging via deep learning. In IEEE 13th International Symposium on Biomedical Imaging (ISBI), Prague, 2016. pp. 514-517.
25. Zhu B, Liu JZ, Cauley SF, Rosen BR, Rosen MS. Image reconstruction by domain-transform manifold learning. Nature 2018;555(7697):487-492.
26. Hammernik K, Klatzer T, Kobler E, Recht MP, Sodickson DK, Pock T, Knoll F. Learning a variational network for reconstruction of accelerated MRI data. Magnetic Resonance in Medicine 2018;79(6):3055-3071.
27. Schlemper J, Caballero J, Hajnal JV, Price AN, Rueckert D. A deep cascade of convolutional neural networks for dynamic MR image reconstruction. IEEE transactions on Medical Imaging 2017;37(2):491-503.
28. Aggarwal HK, Mani MP, Jacob M. MoDL: Model-based deep learning architecture for inverse problems. IEEE transactions on medical imaging 2018;38(2):394-405.
29. Hammernik K, Schlemper J, Qin C, Duan J, Summers RM, Rueckert D. Systematic evaluation of iterative deep neural networks for fast parallel MRI reconstruction with sensitivity-weighted coil combination. Magnetic Resonance in Medicine 2021;86(4):1859-1872.
30. Antun V, Renna F, Poon C, Adcock B, Hansen AC. On instabilities of deep learning in image reconstruction and the potential costs of AI. Proceedings of the National Academy of Sciences 2020;117(48):30088-30095.
31. Guo C, Wu J, Rosenberg JT, Roussel T, Cai S, Cai C. Fast chemical exchange saturation transfer imaging based on PROPELLER acquisition and deep neural network reconstruction. Magnetic Resonance in Medicine 2020;84(6):3192-3205.
32. Ronneberger O, Fischer P, Brox T. U-Net: Convolutional Networks for Biomedical Image Segmentation. In: Proceedings of International Conference on Medical Image Computing and Computer-AssistedIntervention (MICCAI),

Munich, Germany; 2015:234-241.
33. Wang P, Guo P, Lu J, Zhou J, Jiang S, Patel VM. Improving amide proton transfer-weighted mri reconstruction using t2-weighted images. 2020. Springer. p 3-12.
34. Xu J, Zu T, Hsu Y-C, Sun Y, Wu D, Wang X, Zhang Y. Accelerating Chemical Exchange Saturation Transfer Imaging Using a Model-based Deep Neural Network. 2022. arXiv preprint arXiv:2205.10265.
35. Curry HB. The method of steepest descent for non-linear minimization problems. Quarterly of Applied Mathematics 1944;2(3):258-261.
36. Daubechies I. The wavelet transform, time-frequency localization and signal analysis. IEEE transactions on information theory 1990;36(5):961-1005.
37. Block KT, Uecker M, Frahm J. Undersampled radial MRI with multiple coils. Iterative image reconstruction using a total variation constraint. Magnetic Resonance in Medicine: An Official Journal of the International Society for Magnetic Resonance in Medicine 2007;57(6):1086-1098.
38. Sandino CM, Cheng JY, Chen F, Mardani M, Pauly JM, Vasanawala SS. Compressed sensing: From research to clinical practice with deep neural networks: Shortening scan times for magnetic resonance imaging. IEEE signal processing magazine 2020;37(1):117-127.
39. Zhang T, Pauly JM, Levesque IR. Accelerating parameter mapping with a locally low rank constraint. Magnetic resonance in medicine 2015;73(2):655-661.
40. Vishnevskiy V, Walheim J, Kozerke S. Deep variational network for rapid 4D flow MRI reconstruction. Nature Machine Intelligence 2020;2(4):228-235.
41. Huang F, Akao J, Vijayakumar S, Duensing GR, Limkeman M. k‐t GRAPPA: A k‐space implementation for dynamic MRI with high reduction factor. Magnetic Resonance in Medicine 2005;54(5):1172-1184.
42. Tsao J, Boesiger P, Pruessmann KP. k-t BLAST and k-t SENSE: dynamic MRI with high frame rate exploiting spatiotemporal correlations. Magnetic Resonance in Medicine 2003;50(5):1031-1042.
43. Zbontar J, Knoll F, Sriram A, Murrell T, Huang Z, Muckley MJ, Defazio A, Stern R, Johnson P, Bruno M. fastMRI: An open dataset and benchmarks for accelerated MRI. 2018. arXiv preprint arXiv:1811.08839.
44. Knoll F, Zbontar J, Sriram A, Muckley MJ, Bruno M, Defazio A, Parente M, Geras KJ, Katsnelson J, Chandarana H, Zhang Z, Drozdzalv M, Romero A, Rabbat M, Vincent P, Pinkerton J, Wang D, Yakubova N, Owens E, Zitnick CL, Recht MP, Sodickson DK, Lui YW. fastMRI: A Publicly Available Raw k-Space and DICOM Dataset of Knee Images for Accelerated MR Image Reconstruction Using Machine Learning. Radiol Artif Intell 2020;2(1):e190007.
45. Zhou J, van Zijl PCM. Chemical exchange saturation transfer imaging and spectroscopy. Progress in Nuclear Magnetic Resonance Spectroscopy 2006;48(2-3):109-136.
46. Zhang T, Pauly JM, Vasanawala SS, Lustig M. Coil compression for accelerated imaging with Cartesian sampling. Magnetic resonance in medicine 2013;69(2):571-582.
47. Hartigan JA, Wong MA. Algorithm AS 136: A k-means clustering algorithm. Journal of the royal statistical society series c (applied statistics) 1979;28(1):100-108.
48. Likas A, Vlassis N, Verbeek JJ. The global k-means clustering algorithm. Pattern recognition 2003;36(2):451-461.
49. Deng J, Dong W, Socher R, Li L-J, Li K, Fei-Fei L. Imagenet: A large-scale hierarchical image database. In 2009 IEEE conference on computer vision and pattern recognition, 2009. pp. 248-255.
50. Uecker M, Lai P, Murphy MJ, Virtue P, Elad M, Pauly JM, Vasanawala SS, Lustig M. ESPIRiT-An Eigenvalue Approach to Autocalibrating Parallel MRI: Where SENSE Meets GRAPPA. Magnetic Resonance in Medicine 2014;71(3):990-1001.
51. Liu R, Zhang H, Niu W, Lai C, Ding Q, Chen W, Liang S, Zhou J, Wu D, Zhang

51. Y. Improved chemical exchange saturation transfer imaging with real‐time frequency drift correction. Magnetic resonance in medicine 2019;81(5):2915-2923.
52. Liu R, Zhang H, Qian Y, Hsu YC, Fu C, Sun Y, Wu D, Zhang Y. Frequency‐stabilized chemical exchange saturation transfer imaging with real‐time free‐induction‐decay readout. Magnetic Resonance in Medicine 2021;85(3):1322-1334.
53. Zhang Y, Yong X, Liu R, Tang J, Jiang H, Fu C, Wei R, Hsu Y-C, Sun Y, Luo B, Wu D. Whole-brain chemical exchange saturation transfer imaging with optimized turbo spin echo readout. Magnetic Resonance in Medicine 2020;84(3):1161-1172.
54. Kim M, Gillen J, Landman BA, Zhou J, Van Zijl PCM. Water saturation shift referencing (WASSR) for chemical exchange saturation transfer (CEST) experiments. Magnetic Resonance in Medicine 2009;61(6):1441-1450.
55. Kingma DP, Ba J. Adam: A method for stochastic optimization. 2014. arXiv preprint arXiv:1412.6980.
56. Paszke A, Gross S, Massa F, Lerer A, Bradbury J, Chanan G, Killeen T, Lin Z, Gimelshein N, Antiga L. Pytorch: An imperative style, high-performance deep learning library. Advances in neural information processing systems 2019;32.
57. Bhave S, Lingala SG, Johnson CP, Magnotta VA, Jacob M. Accelerated whole‐brain multi‐parameter mapping using blind compressed sensing. Magnetic resonance in medicine 2016;75(3):1175-1186.
58. She H, Greer JS, Zhang S, Li B, Keupp J, Madhuranthakam AJ, Dimitrov IE, Lenkinski RE, Vinogradov E. Accelerating chemical exchange saturation transfer MRI with parallel blind compressed sensing. Magnetic resonance in medicine 2019;81(1):504-513.
59. Zhao H, Gallo O, Frosio I, Kautz J. Loss functions for image restoration with neural networks. IEEE Transactions on computational imaging 2016;3(1):47-57.
60. Wang Z, Bovik AC, Sheikh HR, Simoncelli EP. Image quality assessment: from error visibility to structural similarity. IEEE transactions on image processing 2004;13(4):600-612.

**Figure Captions:**

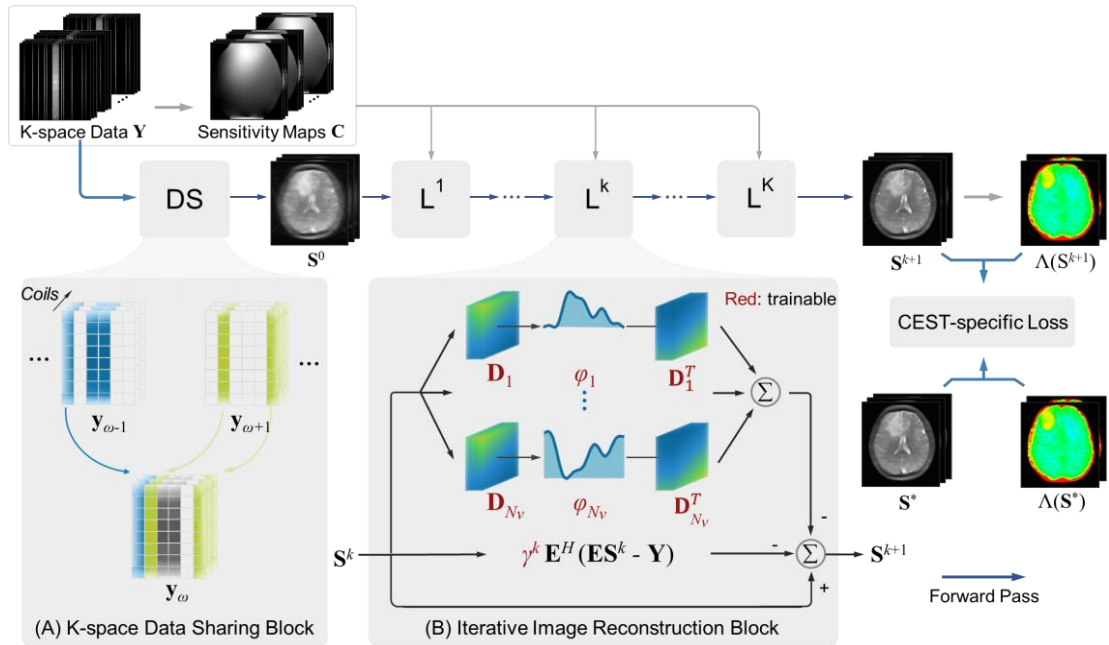

**Figure 1. Illustration of the proposed CEST-VN framework.** CEST-VN is implemented with a *k*-space data sharing (DS) block in *k-ω* space (A), followed by a model-based image reconstruction network with $K$ iterative blocks (B). The CEST-VN framework is input with undersampled *k*-space data and pre-computed sensitivity maps, and outputs the reconstructed source images. A CEST-specific loss function, which jointly measures the errors of CEST source images and CEST contrast maps, is used for network training.

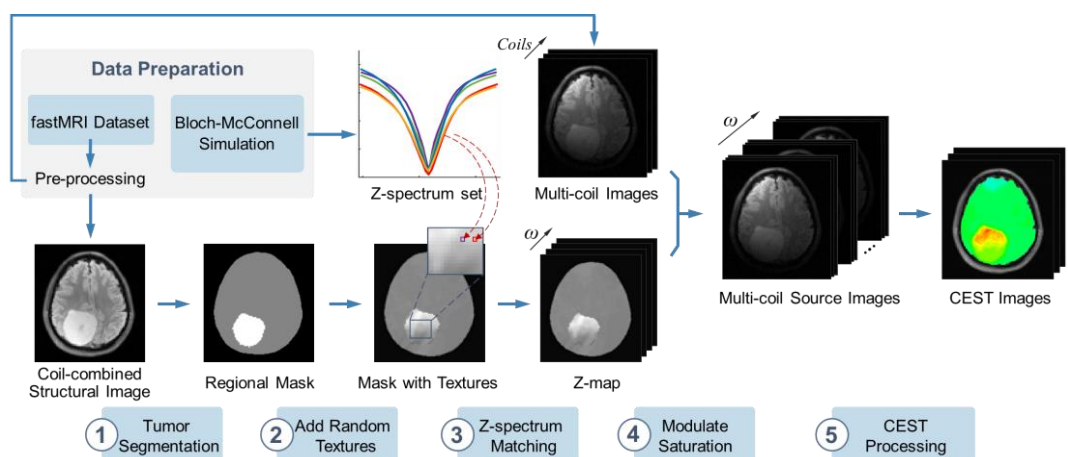

**Figure 2. Flowchart of multi-coil CEST data generation.** A pipeline based on numerical Bloch-McConnell simulations is proposed to simulate multi-coil CEST data from a publicly available anatomical MRI data repository (fastMRI). After initial data preparation, five extra processing steps are implemented, including "Tumor

Segmentation", "Add Random Textures", "Z-spectrum Matching", "Modulate Saturation", and "CEST Processing".

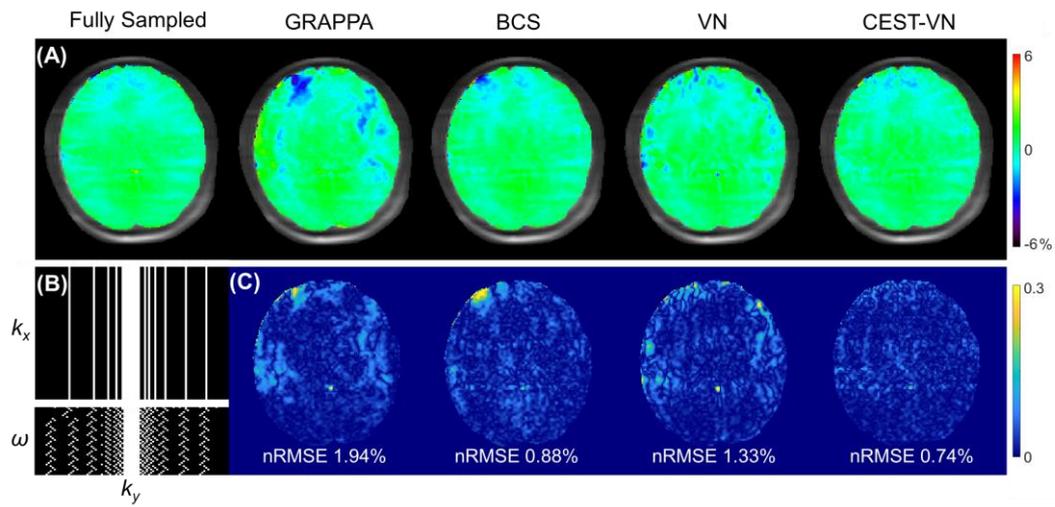

**Figure 3. Comparison of proposed CEST-VN with state-of-the-art methods on a healthy volunteer.** A, Reference fully-sampled APTw map vs. 6-fold-undersampled APTw maps reconstructed by GRAPPA, BCS, original VN, and CEST-VN with the undersampling pattern shown in part (B). B, The 2D $k$-space was variably undersampled in the phase-encoding direction ($k_y$) and fully sampled in the frequency-encoding direction ($k_x$) for each frame. The $k_x$-$k_y$ undersampling pattern was shifted randomly in the $k_y$ direction with the saturation frequency offset ($\omega$). C, Error maps of the accelerated APTw maps against the reference one.

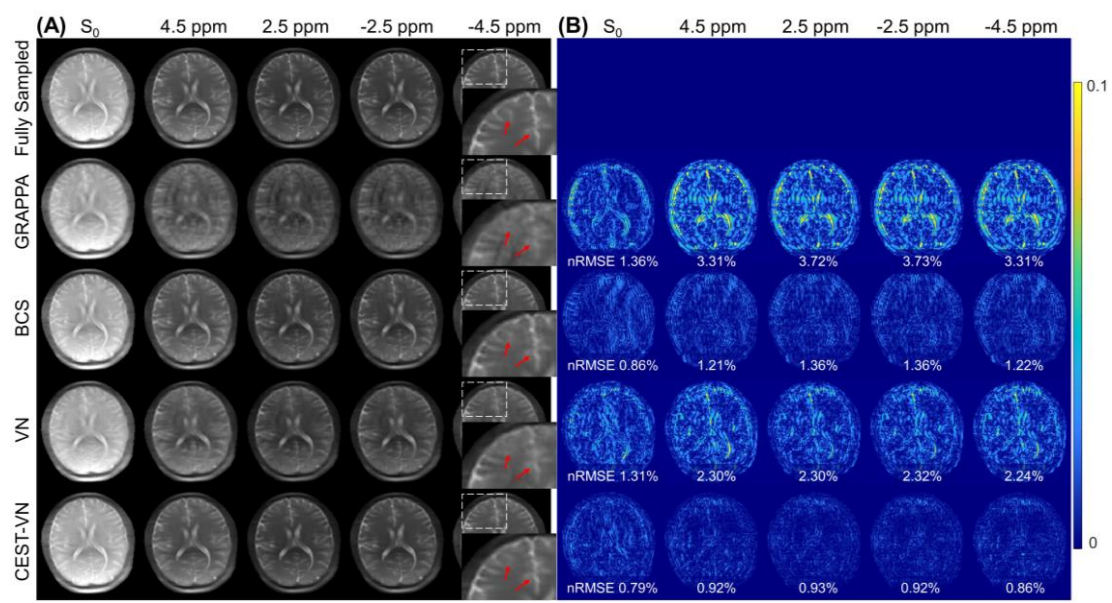

**Figure 4. Reconstructed source images from a healthy volunteer.** A, Unsaturated

images ($S_0$) and saturated images at 4.5, 2.5, -2.5, and -4.5 ppm from different reconstruction methods with an acceleration factor R = 6. B, Error maps of the accelerated source images against the reference fully-sampled ones. The 3D *x-ω* CEST source images were normalized using the maximum absolute value before calculating the error maps.

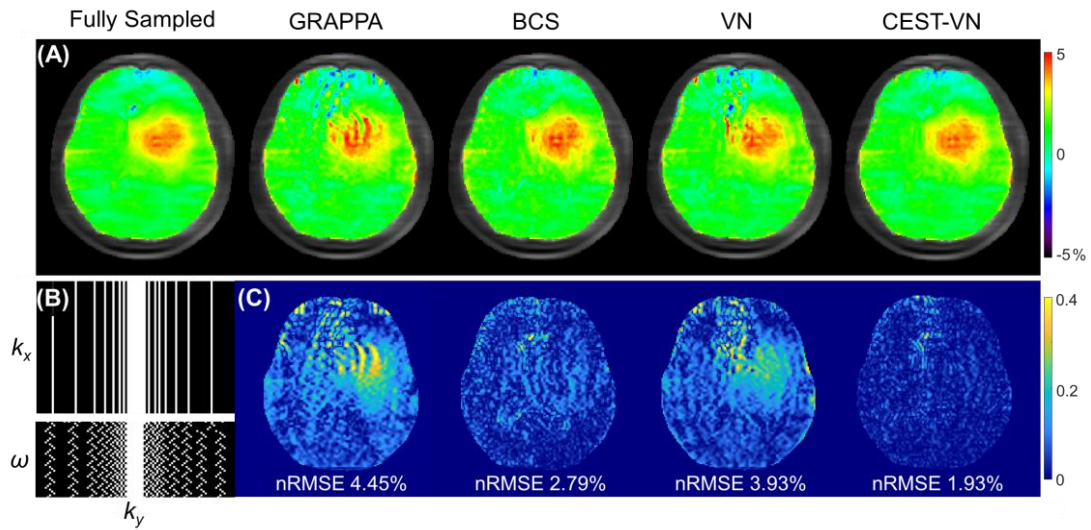

**Figure 5. Comparison of proposed CEST-VN with state-of-the-art approaches on a glioma patient.** A, Reference fully-sampled APTw map vs. 4-fold accelerated APTw maps reconstructed by GRAPPA, BCS, original VN, and CEST-VN with the undersampling pattern shown in part (B). B, The 2D *k*-space was variably undersampled in the phase-encoding direction ($k_y$) and fully sampled in the frequency-encoding direction ($k_x$) for each frame. The $k_x$-$k_y$ undersampling pattern was shifted randomly in the $k_y$ direction with the saturation frequency offset (*ω*). C, Error maps of the accelerated APTw maps against the reference one.

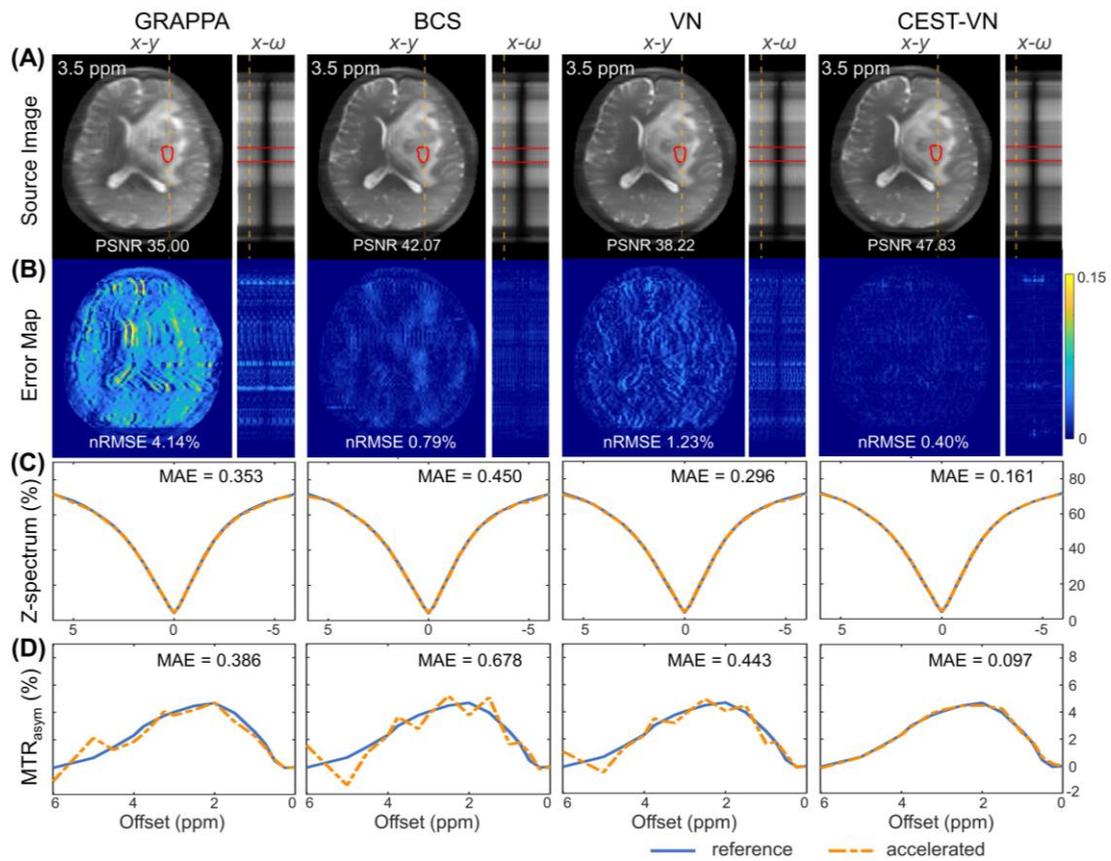

**Figure 6. Reconstructed source images and CEST spectra from a glioma patient.** A-B, Source images (A) reconstructed by different methods and their corresponding error maps (B), with the location of the saturated frame at 3.5 ppm illustrated by yellow dashed lines. C-D, Z-spectra (C) and $MTR_{asym}$ spectra (D) averaged over the region of interest inside the tumor delineated by red curves in part (A).

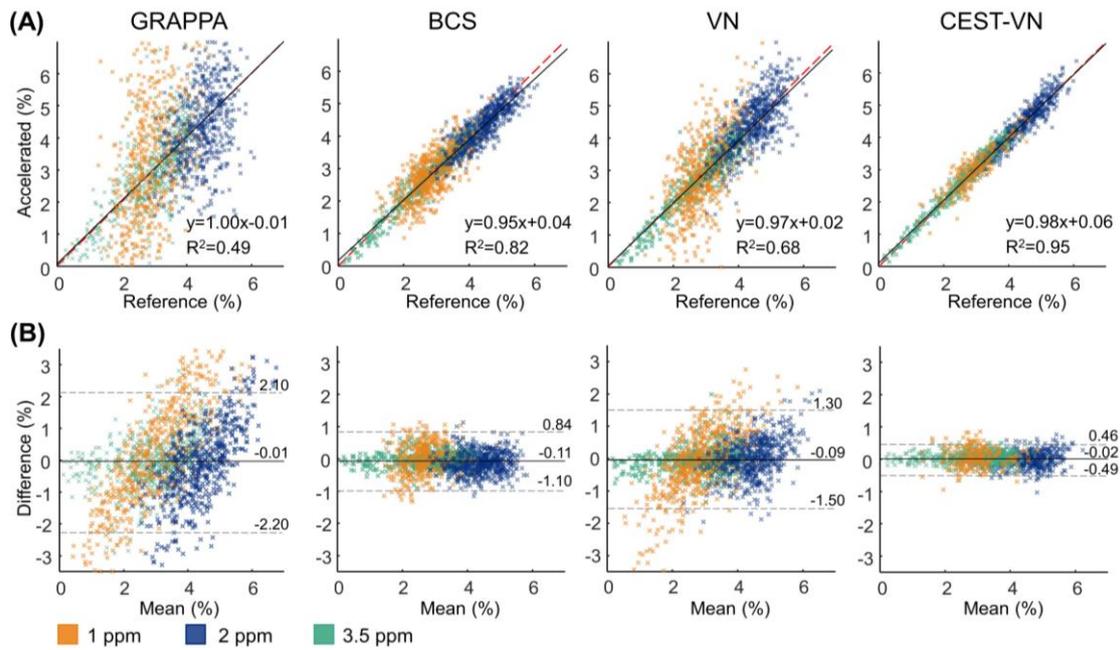

**Figure 7. Quantitative MTR$_{asym}$ results from different reconstruction methods within the whole tumor region.** The pixel-wise correlation analyses (A) and Bland-Altman analyses (B) of MTR$_{asym}$ values generated by the different accelerated methods versus those obtained from the reference fully-sampled data within the whole tumor region of the same patient shown in Figs. 5 and 6.

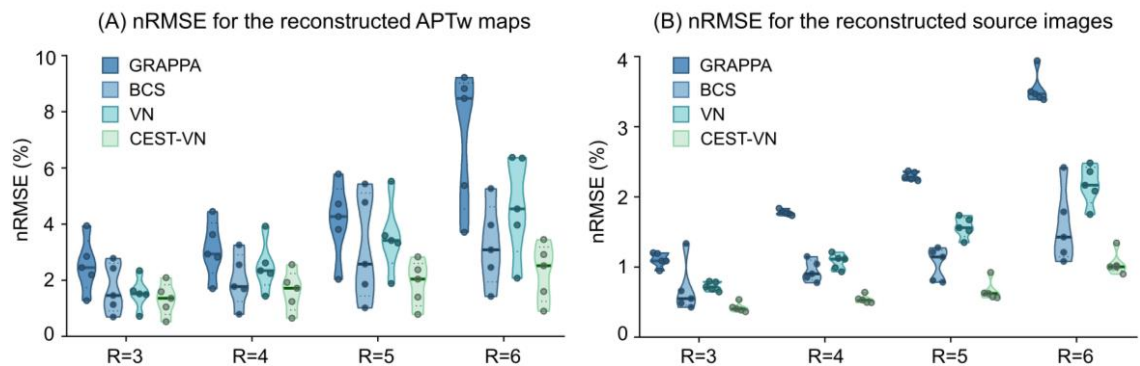

**Figure 8. Violin plots of reconstruction errors on five brain tumor patients.** A, nRMSE values for the APTw maps reconstructed by different methods with varying acceleration factors R from 3 to 6. B, nRMSE values for the source CEST images reconstructed by different methods with varying acceleration factors.

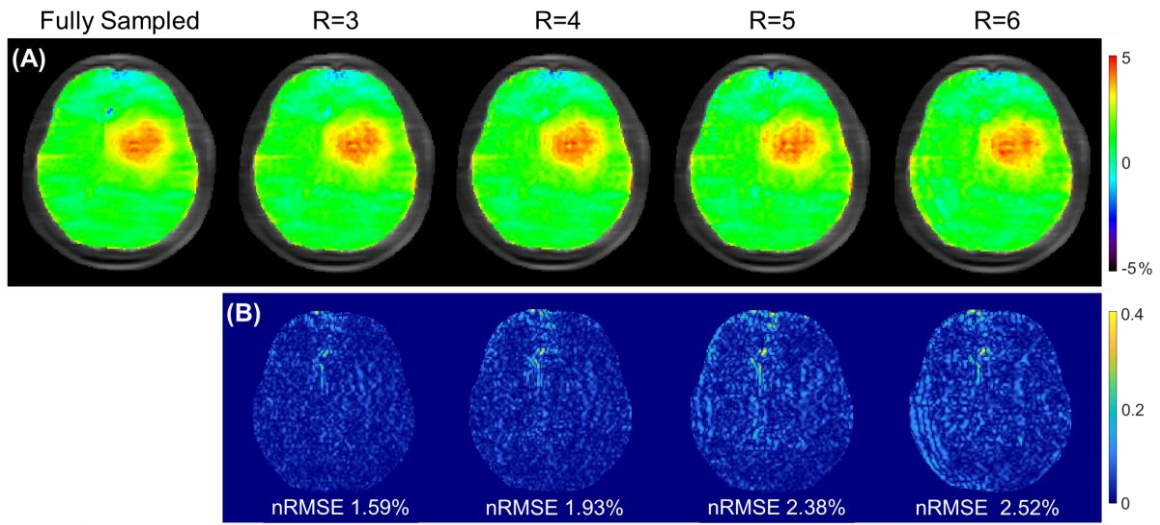

**Figure 9. APTw maps reconstructed by CEST-VN on a brain tumor patient with various acceleration factors.** A, Reference fully-sampled APTw map and accelerated CEST-VN APTw maps with various undersampling factors from 3 to 6. B, Error maps of accelerated APTw maps against the fully-sampled reference one.

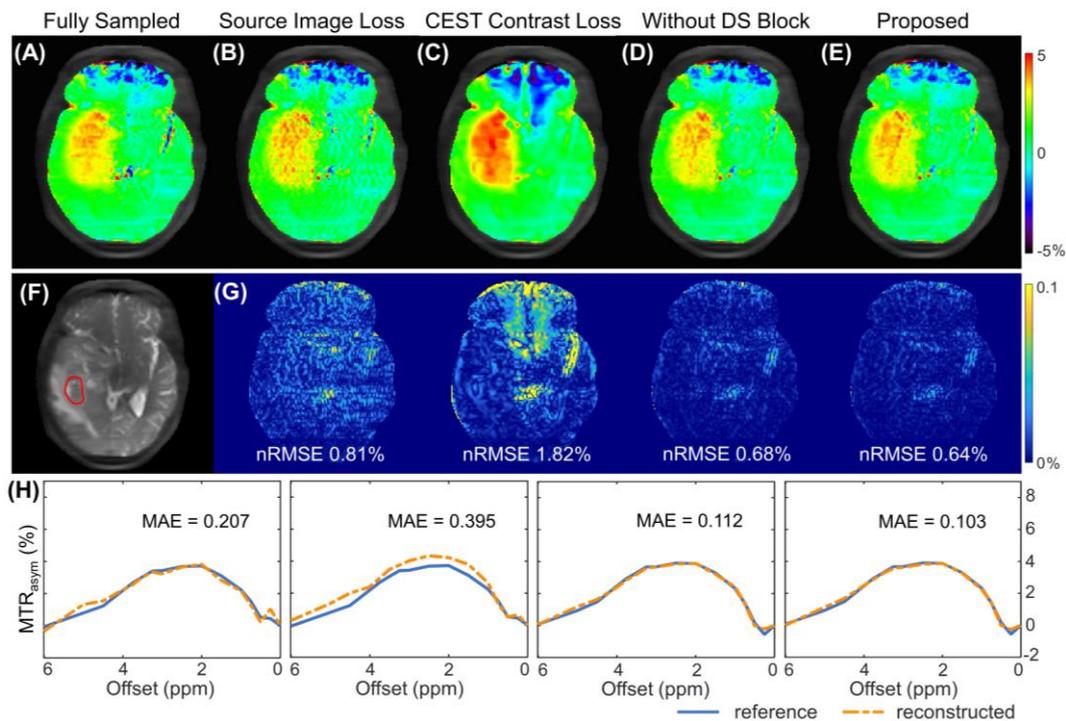

**Figure 10. APTw maps from neural networks trained with different loss functions and settings.** A, Reference fully-sampled APTw maps. B-C, APTw maps yielded by the neural networks trained with only a CEST source image loss (B) or CEST contrast loss (C) function. D, APTw map from the neural network without the data-sharing (DS) block. E, APTw map from the proposed CEST-VN. The corresponding error maps (G)

and MTR$_{asym}$ spectra (H) from the tumor region delineated with red lines in part (F).

# SUPPORTING INFORMATION:

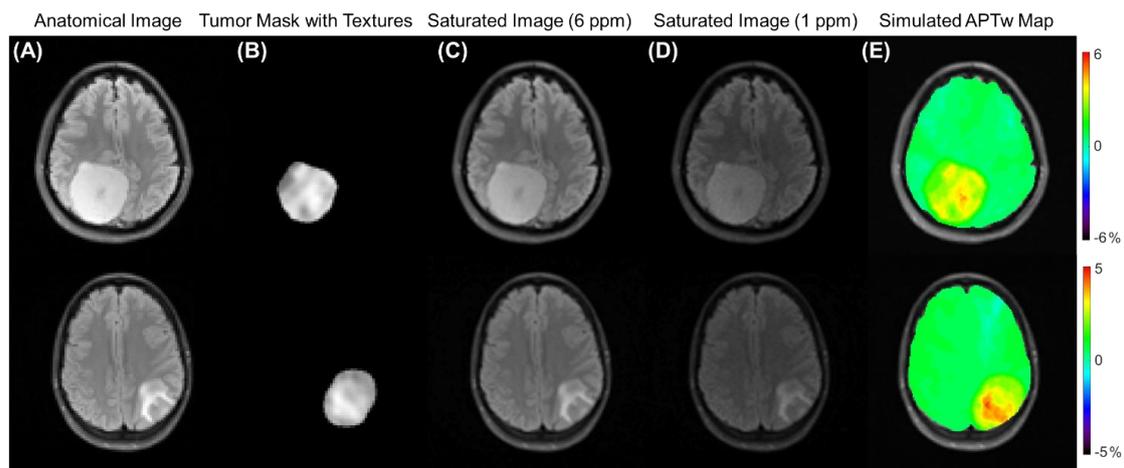

**Figure S1. Results of CEST data simulation.** A, Anatomical images from publicly available fastMRI brain dataset, which were regarded as unsaturated images ($S_0$). B, Segmented or synthesized tumor masks with random textures obtained from natural-scene ImageNet images after smoothing. C-D, Simulated CEST source images at 6 ppm and 1 ppm, respectively. E, APTw maps calculated from the saturated source images.

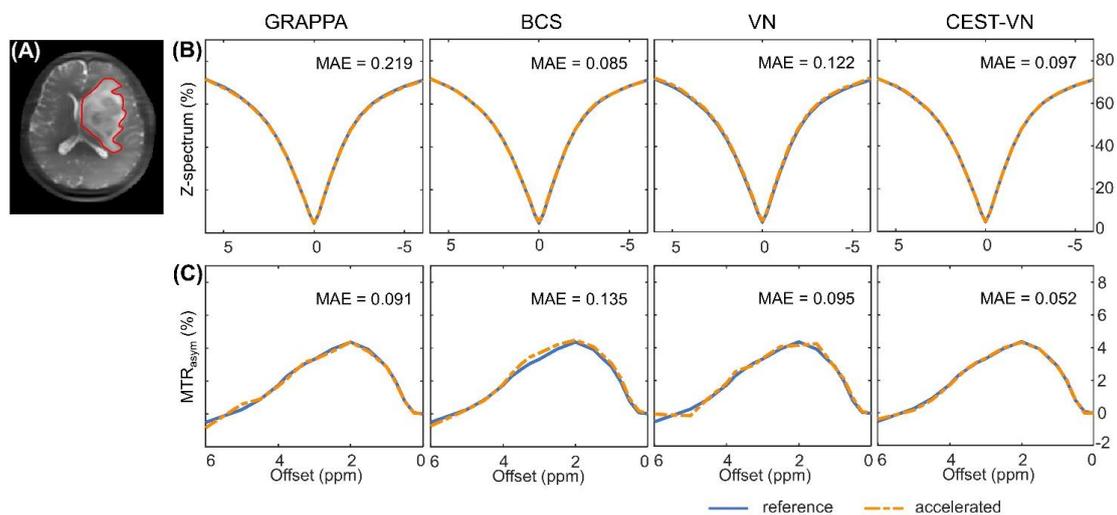

**Figure S2. Reconstructed source image, z-spectra and MTR$_{asym}$ spectra from the glioma patient shown in Fig. 6.** A, Source image illustrating the region of interest chosen (red curve). Z-spectra (B) and MTR$_{asym}$ spectra (C) reconstructed by different acceleration methods compared with the blue reference spectra.

**Table S1. Parameters for the three-pool Bloch-McConnell model used to simulate z-spectra.** Abbreviations: MT, Magnetization transfer; $K_{ws}$, Exchange rate from the water pool to the solute pool (amide or MT).

|  | Tumor | | | Normal | | |
|---|---|---|---|---|---|---|
| **Compound** | Water | Amide | MT | Water | Amide | MT |
| **Δω (ppm)** | 0[1] | 3.5[2] | 0[3] | 0[1] | 3.5[2] | 0[3] |
| **T1 (ms)** | 1400-1700[4,5] | 1000[1] | 1000[1] | 700-1300[3,6] | 1000[1] | 1000[1] |
| **T2 (ms)** | 120-180[4,5] | 2[7] | 0.001[3] | 60-110[6] | 2[7] | 0.001[3] |
| **Concentration (M)** | 74.59[1] | 0.20-0.48[8] | 4.5-7.5[3] | 74.59[1] | 0.01-0.24[8] | 4.5-7.5[3] |
| **Exchange rate $K_{ws}$ (Hz)** |  | 6-12[8] | 3-5[8] |  | 0.3-7.0[8] | 3-5[8] |
| **The number of z-spectra** | 28800 | | | 32400 | | |

**Table S2. nRMSE of reconstructed APTw maps from different methods with acceleration factors of 3–6 in five brain tumor patients.** *P*-values were obtained using paired t-test between CEST-VN and the other accelerated methods.

|  | nRMSE of APTw maps (Mean ± SD%) | | | |
|---|---|---|---|---|
|  | GRAPPA | BCS | VN | CEST-VN |
| R=3 | 2.54 ± 0.87 | 1.70 ± 0.79 | 1.54 ± 0.51 | **1.32 ± 0.53** |
| *P*-value | < 0.05 | 0.146 | 0.379 |  |
| R=4 | 3.10 ± 0.91 | 2.02 ± 0.84 | 2.51 ± 0.81 | **1.62 ± 0.64** |
| *P*-value | < 0.05 | < 0.05 | < 0.05 |  |
| R=5 | 4.12 ± 1.23 | 3.10 ± 1.74 | 3.55 ± 1.16 | **1.89 ± 0.72** |
| *P*-value | < 0.05 | 0.087 | < 0.05 |  |
| R=6 | 7.13 ± 2.18 | 3.16 ± 1.44 | 4.67 ± 1.61 | **2.28 ± 0.92** |
| *P*-value | < 0.05 | 0.087 | < 0.05 |  |

**Table S3. nRMSE of reconstructed CEST source images from different methods with acceleration factors of 3–6 in five brain tumor patients.** nRMSE was calculated from CEST source images at all saturation and unsaturation offsets. *P*-values were obtained using paired t-test between CEST-VN and the other accelerated methods.

|  | nRMSE of source images (Mean ± SD%) | | | |
| --- | --- | --- | --- | --- |
|  | GRAPPA | BCS | VN | CEST-VN |
| R=3 | 1.11 ± 0.09 | 0.69 ± 0.33 | 0.73 ± 0.05 | **0.42 ± 0.06** |
| *P*-value | < 0.05 | 0.125 | < 0.05 |  |
| R=4 | 1.78 ± 0.04 | 0.95 ± 0.13 | 1.08 ± 0.10 | **0.54 ± 0.05** |
| *P*-value | < 0.05 | < 0.05 | < 0.05 |  |
| R=5 | 2.30 ± 0.05 | 1.05 ± 0.21 | 1.57 ± 0.13 | **0.69 ± 0.19** |
| *P*-value | < 0.05 | 0.119 | < 0.05 |  |
| R=6 | 3.60 ± 0.31 | 1.46 ± 0.58 | 2.17 ± 0.25 | **1.05 ± 0.15** |
| *P*-value | < 0.05 | 0.173 | < 0.05 |  |

**Reference**

1. Cai K, Haris M, Singh A, Kogan F, Greenberg JH, Hariharan H, Detre JA, Reddy R. Magnetic resonance imaging of glutamate. Nature medicine 2012;18(2):302-306.
2. Zhou J, Lal B, Wilson DA, Laterra J, Van Zijl PCM. Amide proton transfer (APT) contrast for imaging of brain tumors. Magnetic Resonance in Medicine 2003;50(6):1120-1126.
3. van Zijl PCM, Lam WW, Xu J, Knutsson L, Stanisz GJ. Magnetization Transfer Contrast and Chemical Exchange Saturation Transfer MRI. Features and analysis of the field-dependent saturation spectrum. Neuroimage 2018;168:222-241.
4. Zhang Y, Liu X, Zhou J, Bottomley PA. Ultrafast compartmentalized relaxation time mapping with linear algebraic modeling. Magnetic resonance in medicine 2018;79(1):286-297.
5. Badve C, Yu A, Dastmalchian S, Rogers M, Ma D, Jiang Y, Margevicius S, Pahwa S, Lu Z, Schluchter M. MR fingerprinting of adult brain tumors: initial experience. American Journal of Neuroradiology 2017;38(3):492-499.
6. Stanisz GJ, Odrobina EE, Pun J, Escaravage M, Graham SJ, Bronskill MJ, Henkelman RM. T1, T2 relaxation and magnetization transfer in tissue at 3T. Magnetic Resonance in Medicine 2005;54(3):507-512.
7. Zhang XY, Wang F, Xu J, Gochberg DF, Gore JC, Zu Z. Increased CEST specificity for amide and fast-exchanging amine protons using exchange-dependent relaxation rate. NMR in Biomedicine 2018;31(2):e3863.
8. Wen Q, Wang K, Hsu YC, Xu Y, Sun Y, Wu D, Zhang Y. Chemical exchange saturation transfer imaging for epilepsy secondary to tuberous sclerosis complex at 3T: Optimization and analysis. NMR in Biomedicine 2021;34(9):e4563.